\documentclass[
reprint,
superscriptaddress,
 amsmath,amssymb,
 aps,
]{revtex4-2}

\usepackage{graphicx}
\usepackage{dcolumn}
\usepackage{bm}

\begin{document}

\preprint{APS/123-QED}

\title{Measurements and modeling of swimming speed dependence on stroke frequency in scyphozoan jellyfish}

\author{Noa K. Yoder}
\affiliation{Graduate Aerospace Laboratories, California Institute of Technology, Pasadena, CA 91125, USA}

\author{John O. Dabiri}
\affiliation{Graduate Aerospace Laboratories, California Institute of Technology, Pasadena, CA 91125, USA}
\affiliation{Mechanical and Civil Engineering, California Institute of Technology, Pasadena, CA 91125, USA}

\date{\today}

\begin{abstract}
Scyphozoan jellyfish exhibit the highest locomotive efficiency in the animal kingdom, making them of particular interest in fluid dynamics and bioinspired robotics. Despite this, prevalent analytical models of jellyfish swimming have been based on the swimming traits of hydrozoan jellyfish which utilize jet propulsion, rather than scyphozoan jellyfish which utilize paddling propulsive methods. Additionally, while stroke frequency is a driving variable in speeds achieved by undulatory swimmers, a similar dependence has not been previously explored for jellyfish. This work investigates the relationship between stroke frequency and swimming speeds in two species of scyphozoan jellyfish, \textit{Aurelia aurita} and \textit{Cassiopea xamachana}. An experimental study was conducted using a biohybrid technique that controls the muscle contraction frequency of freely swimming, live jellyfish with portable, implanted microelectronics. Swimming speeds were measured from video recordings in a 2.4 meter tall water tank. It was found that despite differences in their natural swimming frequencies, the \textit{Aurelia} and \textit{Cassiopea} displayed similar speed-frequency relationships with peak swimming speeds occurring at $0.55\pm0.05$ Hz and $0.50\pm0.05$ Hz respectively. The difference in natural stroke frequency displayed by scyphomedusea despite the shared relationship between swimming speed and stroke frequency in these two species, suggests that natural stroke frequency may be more related to other functions such as filter feeding, rather than locomotion. A new analytical model developed for scyphozoan, paddling jellyfish was shown to have closer agreement with the experimental results than existing models based on jet propulsion. The model demonstrated the driving factors in the relationship between swimming speed and stroke frequency to be the speed of the jellyfish bell margin and changes in body kinematics with stroke frequency.
\end{abstract}


\maketitle


\section{Introduction}

The most common method of propulsion in aquatic vertebrates is undulatory swimming \citep{sanchez-rodriguez_scaling_2023}, wherein the animal passes a wave of lateral displacement from its anterior to posterior end. This motion causes the formation and shedding of vortices on either side of the animal, and the rearward momentum imparted to the surrounding fluid propels the animal forward \citep{lauder_hydrodynamics_2005}. Swimming speeds achieved by this method of locomotion have been shown to depend primarily on the frequency and lateral amplitude of the swimming stroke. Undulatory swimming speeds are found to be positively correlated with oscillation frequency \citep{bainbridge_speed_1958, videler_fish_1991, sanchez-rodriguez_scaling_2023} due to more frequent transfers of rearward momentum to the surrounding fluid.

Invertebrates, on the other hand, exhibit a variety of other locomotive methods including jet propulsion \citep{gemmell_cool_2021}, paddling propulsion, and the coordinated movement of many cilia \citep{chia_locomotion_1984}. Here we focus on the swimming methods of scyphomedusae due to their high energetic efficiency \citep{gemmell_passive_2013}. Their combination of low energy expenditure and simple morphology make scyphozoan jellyfish a focus of many bio-inspired robotic devices \citep{ cheng_untethered_2018, villanueva_biomimetic_2011, wang_versatile_2023}.

\textit{Aurelia aurita}, or moon jellies, in particular, have been used as a representative organism in a variety of studies due to their global availability \citep{costello_hydrodynamics_2021} and wide range of relevant Reynolds numbers at which they are observed to swim throughout their lifecycles, i.e. $1<Re<10^4$ \citep{feitl_functional_2009}. These studies found that naturally swimming, mature, \textit{Aurelia} exhibit a mean peak frequency value of approximately $0.24\pm0.11$ Hz \citep{xu_low-power_2020}.

\textit{Cassiopea} jellyfish, a species that does not naturally swim as often as \textit{Aurelia}, provide an interesting contrast. Both species belong to the scyphozoan class and in their mature life stage posses an oblate form factor (low fineness ratio of bell height to bell diameter). Both also utilize a paddling swimming stroke. Despite these similarities, \textit{Aurelia} and \textit{Cassiopea} display extremely disparate behaviors and natural stroke frequencies. Unlike the open water \textit{Aurelia}, which utilize low stroke frequencies, \textit{Cassiopea} are benthic, residing on the sea floor in shallow waters. They utilize higher average pulse frequencies of approximately $0.96\pm0.26$ Hz during the day and $0.65\pm0.17$ Hz at night \citep{nath_jellyfish_2017}.

\begin{figure*}[t!]
\centering\includegraphics[width=5.4in]{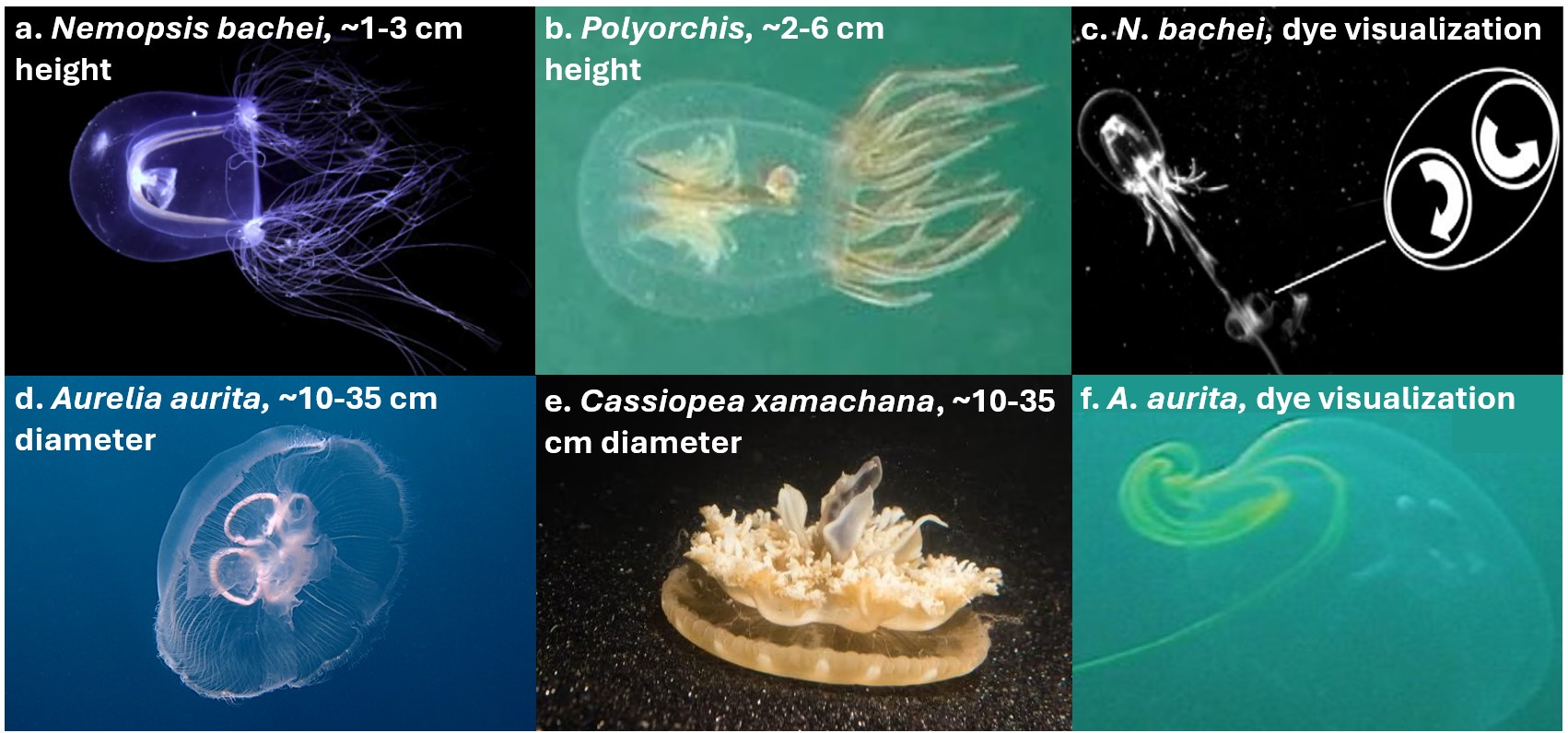}
\caption{\textbf{Characteristic examples of hydrozoan and scyphozoan jellyfish.} The top images, depict hydrozoan jellyfish, the \textit{Nemopsis bachei} (a, c) and \textit{Polyorchis} (b) \citep{jellyfishwarehouse_crosshair, seanet_hydrozoa}. These were two of the jellyfish used in developing previous analytical models of jellyfish swimming  \citep{dabiri_fast-swimming_2006,daniel_mechanics_1983,gladfelter_comparative_1973}. The bottom images, depict the \textit{Aurelia aurita} (d, f) \citep{noauthor_european_nodate} and \textit{Cassiopea xamachana} (e) \citep{association_caught_2017}, scyphozoan jellyfish species used in this work. The hydrozoan jellyfish tend to be more prolate, while the scyphozoan jellyfish more oblate. The right most images show dye visualizations from previous studies of the respective wakes \citep{dabiri_fast-swimming_2006, dabiri_flow_2005}. The hydozoan jellyfish display a vortex ring at the end of a jet, while the scyphozoan jellyfish shows no jet, only the vortex ring.}
\label{species}
\end{figure*}

The aforementioned observations of a range of naturally occurring stroke frequencies in jellyfish with similar body shapes and swimming kinematics raise the question of how those differences affect the swimming hydrodynamics. Previous analytical descriptions of jellyfish swimming predict that swimming speed should increase with stroke frequency until constrained by the body's physical limitations. These studies utilized kinematics-based momentum models \citep{daniel_mechanics_1983} and vortex wake models \citep{dabiri_fast-swimming_2006} to determine swimming speed.

In both models the thrust depends on the volume and velocity of the fluid exiting the jellyfish subumbrellar cavity, which is determined by the change in the subumbrellar volume. Assuming the kinematics of a stroke are independent of the frequency at which the stroke occurs, then the impulse imparted per stroke, I, remains approximately constant. At steady state swimming the time averaged thrust $\bar{T} = If$ can be balanced with the quasi-steady drag $D=\frac{1}{2}\rho C_d A \bar{U}^2$. This leads to the approximation that the average swimming speed, $\bar{U}$ should grow as $f^{1/2}$. At higher frequencies, it can be predicted that the assumption of frequency-independent stroke kinematics will break down. Reduced stroke amplitudes will lead to a deviation from the $f^{1/2}$ scaling. These models, however, have been based on the swimming strokes of hydrozoan jellyfish which swim using jet-propulsion. A key objective of this work was to determine the applicability of these models to scyphozoan jellyfish, and to develop an alternative scyphomedusea-based model that better estimated vortex volume and circulation for their paddling swimming stroke.

Hydrozoan jellyfish (figure 1a-c), tend to be more prolate, and use a velar muscle around the inner bell margin to assist in ejecting fluid in jet propulsion. By contrast, scyphozoan jellyfish (figure 1d-f), tend to be more oblate, leading to a paddling propulsion method. Previous work has identified interactions between vortices formed during bell contraction and relaxation, respectively, as being key to efficient swimming in scyphozoan jellyfish \citep{costello_hydrodynamics_2021}. Moreover, passive recapture of energy via interactions between the wake vortices and the subumbrellar surface of the animal has been shown to increase swimming distance by 30\% in each swim cycle \citep{gemmell_passive_2013}. The dependence of these different fluid dynamic phenomena on swimming stroke frequency has not previously been studied.

Previous experimental studies of the relationship between stroke frequency and swimming speeds in fish have made use of their optomotor reflex \citep{plaut_critical_2001, videler_fish_1991}--a tendency of fish to hold position in oncoming current. That reflex allows swimming speed to be varied experimentally by increasing oncoming flow, either through a flume or a circular rotating tank. Jellyfish, however, do not have this same response, requiring a new experimental approach.

Recent work has demonstrated that jellyfish swimming stroke frequency can be controlled in freely swimming animals by implanted microelectronics that electrically stimulate the subumbrellar muscle sheet \citep{xu_low-power_2020, xu_field_2020}. Electric signals transmitted through electrodes embedded in the jellyfish bell muscle cause the muscle to contract and the jellyfish to swim. The jellyfish tested in these studies do not have nociceptors or central nervous systems, indicating they have little potential to experience pain \citep{xu_ethics_2025}. Lab and field testing of these devices have demonstrated a nearly three-fold increase in swimming speed compared to uncontrolled swimming of the same distance. The increased swimming speed could be due to stimulation closer to an optimal frequency of actuation, or due to more consistent swimming over long periods of time. While that prior work demonstrated a range of stroke frequencies in which the swimming stroke could be controlled, from approximately 0.3 Hz to 0.8 Hz, the measurements were inconclusive regarding a trend in swimming speed with swimming frequency.

Here we used the biohybrid technique to conduct a comprehensive study of the effects of stroke frequency on swimming speeds in scyphozoan jellyfish. Stroke frequency was systematically varied in two species of scyphozoan jellyfish and their swimming speeds were quantified. These results were then compared to previously developed analytical models of hydromedusean swimming as well as a new model of scyphomedusean swimming. The new model used different parameters that are more relevant in oblate, paddling jellyfish in order to calculate vortex volume, circulation and thrust.

\section{Materials and methods}

\subsection{Animal husbandry}
Adult \textit{Aurelia aurita} ranging in size from 12 to 14 cm were collected from the Cabrillo Marine Aquarium (San Pedro, CA, USA) for laboratory experiments. Between experiments the jellyfish were housed in a 453 liter pseudokreisel tank (Jelliquarium 360, Midwater Systems, Thousand Oaks, CA, USA) filled with 36 ppt artificial seawater and kept at 21$^{\circ}$ C. The jellyfish were fed twice daily with live \textit{Artemia franciscana} brine shrimp, occasionally supplemented with dry food made from ground plankton (Jellyfish Art, Fort Lauderdale, FL, USA).

\textit{Cassiopea xamachana} were procured from Dynasty Marine Associates, Inc (Marathon, FL, USA). This species of jellyfish is benthic and does not require circulating currents to simulate the open ocean. Therefore between experiments they were housed in two rectangular tanks with quiescent seawater. The tank water used was the same as the pseudokreisel, but allowed to reach higher temperatures of 23$^{\circ}$ C consistent with their natural, shallow-water habitats. The jellyfish were fed once a day (twice when used in experiments) with frozen capsules of \textit{Artemia franciscana} brine shrimp.

\textit{Cassiopea} have a higher natural pulse frequency than \textit{Aurelia}, which has been found to interfere with the external stimulation of the bell muscle \citep{anuszczyk_increasing_2025}. In order to avoid endogenous stimulation of swimming, the swimming pacemakers of these jellyfish (i.e. rhopalia) were dissected out of the \textit{Cassiopea} before the experiments were conducted. Jellyfish regrow these nerve clusters within one to two days, and one \textit{Cassiopea} was kept under observation until it regenerated enough rhopalia to initiate endrogenous contractions once again.

Both \textit{Aurelia} and \textit{Cassiopea} jellyfish belong to the scyphozoan class which have the most diffuse nervous system of the jellyfish family. In addition, jellyfish do not have nociceptors indicating that it is unlikely they experience pain. An ethical case study of their use in biohybrid experiments found that jellyfish equipped with this technology continue to thrive and even reproduce after it has been removed \citep{xu_ethics_2025}. Nonetheless, in accordance with the reduction principle espoused in that work, a minimal number of test subjects necessary to draw scientific conclusions were used. Experiments were conducted with two \textit{Aurelia} with body diameters of 12.0 $\pm$ 0.5 cm and 14.0 $\pm$ 0.5 cm as well as with three \textit{Cassiopea} with body diameters of 13.0 $\pm$ 0.5 cm, 14.0 $\pm$ 0.5 cm, and 15.0 $\pm$ 0.5 cm.

\subsection{Stroke frequency robotic control}
Control of swimming frequency was achieved by implanting a portable microcontroller in the jellyfish as in previous work \citep{xu_low-power_2020}. The swim controller was housed in an O-ring sealed custom housing 3D printed from photopolymer resin. It contained a 14 mm diameter microcontroller (TinyLily with Atmel Atmega328P processor, TinyCircuits, Akron, OH, USA) and a nominally 3.7 V, 15 mAh, lithium polymer battery cell (PowerStream Technology Inc. Orem, UT, USA).  Perfluoroalkoxy-coated wires passed through the housing (sealed with epoxy around the penetrations) and transmitted an electric signal to two electrodes embedded in the jellyfish bell muscle. The signal was a square wave with 10 ms intervals of battery voltage followed by 0 V intervals whose length depended on the stroke frequency desired. The electrodes were composed of light emitting diodes (LEDs) that allowed for visual confirmation of the electric signal and platinum rod tips that were embedded in the jellyfish muscle with minimal irritation to the tissue. The swim controller housing was positioned on the subumbrellar side and secured with a wooden pin through the jellyfish center. A weighted stopper was attached on the exumbrellar side to provide ballast and keep the pin in place.
\begin{figure}[h]
\includegraphics[width=3.5in]{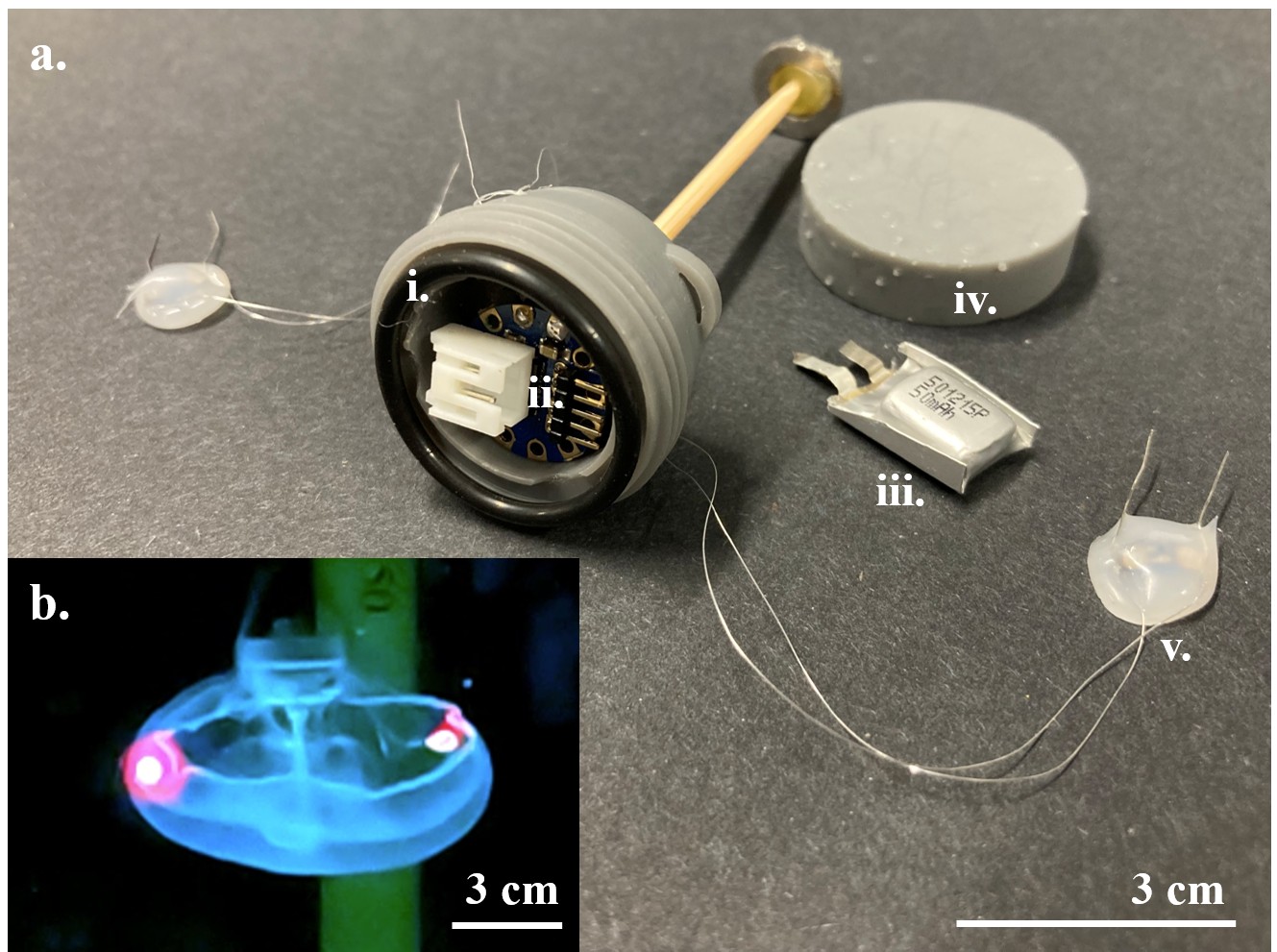}
\caption{\textbf{Biohybrid jellyfish swim controller.} (a) The swim controller is seen opened, revealing the o-ring seal (i.) and TinyLily microcontroller (ii). The unconnected battery (iii.) and lid (iv.) sit to the side. The electrodes (v.) consisting of LEDs and platinum rods are connected to the microcontroller through penetrations in the side of the housing. (b) The swim controller is seen embedded in a live jellyfish during frequency variation experiments, with stimulating electrodes illuminated.}
\label{components}
\end{figure}
\subsection{Swimming performance measurements}
Swimming performance measurements were conducted in a 2.4 m tall test tank, maintained at 36 ppt salinity to match the salinity of the pseudokreisel and \textit{Cassiopea} tanks. The swim controller was ballasted to be slightly positively buoyant, within 0.1 g, to ensure that the swimming achieved was based on solely the jellyfish propulsion and not passive sinking. The swim controller was programmed to cycle through nine different frequencies, $f=[0.30, 0.40, 0.45, 0.50, 0.55, 0.60, 0.65, 0.70, 0.80]$ Hz, and a no stimulation rest period at intervals of 4 min for each frequency. The starting frequency was not pre-selected and the frequencies were cycled multiple times with short intervals, rather than once with long intervals, to prevent biases introduced by swimming fatigue or battery depletion toward the end of the experiment. The frequencies were chosen based on \textit{Aurelia} testing that revealed above 0.80 Hz it was difficult for the jellyfish to respond to stimulus, and below 0.30 Hz, they did not make forward progress with the positively buoyant controller.

The tested frequencies corresponded with Strouhal numbers ($St=\frac{fD}{U}$) between 5 and 27. Previous work has demonstrated that bony fish swimming tends to fall in a significantly lower range of $0.2<St<0.4$ \citep{taylor_flying_2003}. Oblate jellyfish, however, typically swim at $St> 1$ \citep{mchenry_ontogenetic_2003, herschlag_reynolds_2011}. The larger Strouhal numbers in this study are due to the slower swimming speeds caused by the positive buoyancy of the controller.

\begin{figure}[h]
\centering\includegraphics[width=3in]{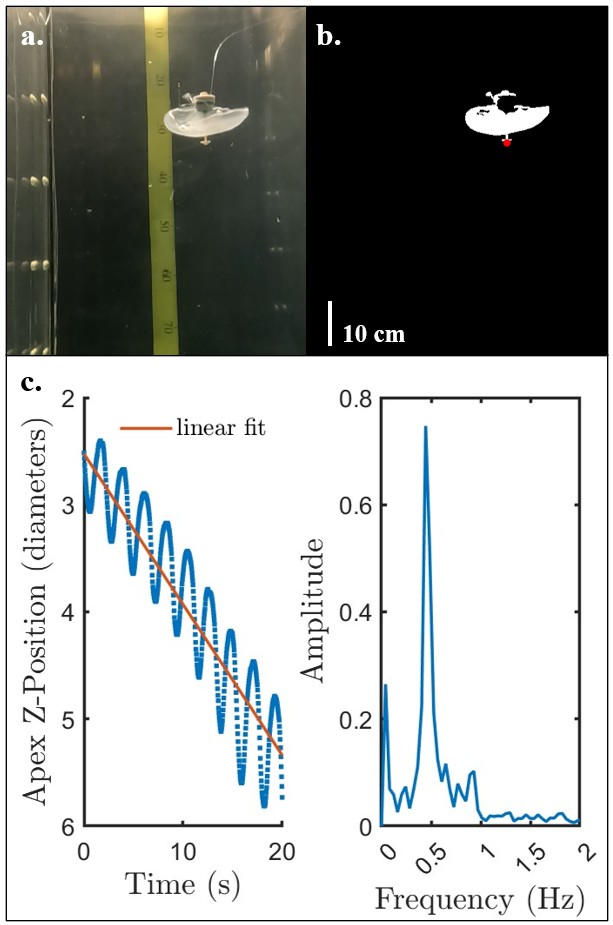}
\caption{\textbf{Data extraction from the swimming performance videos.} A frame from the frequency variation videos (a) depicts the jellyfish equipped with swim controller being stimulated at 0.45 Hz. That same video frame has been converted to a black and white image (b) and white areas are filtered leaving only the jellyfish pixels. The apex of the jellyfish area is identified in red. The apex is tracked across multiple frames providing the plot of jellyfish depth versus time (c). The depth is normalized by jellyfish diameter--also extracted from the video frames. Individual pulses are clearly identifiable in the plot with an overall downward trajectory. The linear fit used to extract the average velocity is plotted over the instantaneous position. The power spectral density of the track (c) confirms that the jellyfish stimulation frequency lies between 0.4 and 0.5 Hz. }
\label{tracking}
\end{figure}

Jellyfish were released approximately 1 body diameter from the top of the tank and swam down against their positive buoyancy until they reached the bottom edge of the camera view, approximately 7 body diameters from the top of the tank for the \textit{Aurelia} experiments, and 15 body diameters for the \textit{Cassiopea} experiments. They were then retrieved and returned to their starting position using a 0.3 mm diameter polymer slack line attached to the housing. Video was taken using a 1080 by 1920 pixel CMOS camera (GoPro Hero10, San Mateo, CA, USA) recording at 30 frames per second.

Video was processed in Matlab by converting each frame to a black and white image. An automated script identified white areas and filtered them by size to select for only the jellyfish pixels. The apex of the exumbrellar surface was then recorded for each frame providing a track of jellyfish position in time. The width of the jellyfish in pixels was also recorded for every frame and a windowed average over 10 s was used to normalize the jellyfish position by body diameter. This provided the dual purpose of normalizing between the different test subjects, and accounting for variation in animal image size introduced by jellyfish drift normal to the plane of the camera. 

The vertical position of the exumbrellar apex oscillated in time due to the unsteady pulsing motion of the jellyfish. Contraction frequencies were confirmed by calculating the power spectral density of the jellyfish position. A least squares estimation of linear best fit was calculated for each track of apex position (excluding the initial acceleration period of at least 10s). The slope of the linear best fit represents the average swimming speed of the jellyfish for that 
frequency.

\subsection{Analytical swimming models}
Three analytical models were investigated to interpret the empirical measurements. All three models calculated the jellyfish velocity by iteratively solving a force balance, based in prior work on jellyfish energetics by Daniel \citep{daniel_mechanics_1983}. The balance includes the jellyfish thrust (T), and the quasi-steady drag (D), versus the mass ($m_{jelly}$), added mass ($\alpha$) and jellyfish acceleration ($\frac{dU_{jelly}}{dt}$). The present extension of this model also included buoyancy (B) to better describe the positively buoyant biohybrid jellyfish.

\begin{align}
T + B + D &= (1 + \alpha) \, m_{\text{jelly}} \, \frac{dU_{\text{jelly}}}{dt}
\end{align}

In all three models the jellyfish was simplified to a hemielipsoid with quasi-steady drag based on the time-dependent speeds and cross-sectional areas of the jellyfish. The coefficient of drag was taken to be that of a hemielipsoid (i.e. $C_d=0.42$), when moving forward and that of a flat plate (i.e. $C_d=1.17$)\citep{sighard_f_hoerner_fluid-dynamic_1965} when moving backward, which occured for some buoyancies. The drag term in all models was given as follows.

\begin{align}
    D &= \frac{1}{2} \rho_w C_d U_{\text{jelly}}^2 A_{\text{jelly}}
\end{align}

Changes in body diameter due to muscle contractions were modeled as a piecewise sinusoidal function. These piecewise sinusoids were used to compute the jet ejection speed from the change in subumbrellar volume \citep{xu_low-power_2020}. The minimum and maximum diameters, as well as the contraction and relaxation periods were estimated based on empirical measurements described in the next section.

The differences in the three models arose from the method of calculating thrust. The first thrust equation ($T_1$) was adapted from a model for jet speed based on conservation of mass \citep{daniel_mechanics_1983}. Here, the thrust generated by the jellyfish was quantified using the mass flow rate and velocity of fluid ejected from the bell which are both functions of the change in subumbrellar volume, $V_{\text{sub}}$.
\begin{align}
    T_1 &= \frac{\rho_w}{A} \left( \frac{dV_{\text{sub}}}{dt} \right)^2
\end{align}

The second approach to calculating thrust ($T_2$) was adapted from a vortex thrust model originally developed and compared against hydromedusan jellyfish \citep{dabiri_fast-swimming_2006}. Here the body kinematics which provided the subumbrellar volume were used to estimate the volume of the vortex expelled $V_v$, including an assumed 30\% entrainment term.
\begin{align}
V_{v,2} &= \frac{1}{1 - 0.3} \, V_{\text{sub}}
\end{align}
This vortex volume could then be used to approximate the radius of the vortex ring, $R$.
\begin{align}
R_2 &= \left( \frac{3}{4\pi} \, V_{v,2} \right)^{1/3}
\end{align}
The circulation in the vortex ring was approximated using a slug model that compares the velocity of the fluid expelled from the jellyfish, $U_\text{jet}$, to the velocity of the surrounding fluid, e.g. the velocity of the jellyfish, $U_\text{jelly}$.
\begin{align}
\Gamma_2 &= \frac{1}{2} \int \left( U_{\text{jet}}^2 - U_{\text{jelly}}^2 \right) \, dt
\end{align}

The velocity of the vortex ring, $U_v$ could then be estimated by a ratio of the vortex circulation to the vortex radius.

\begin{align}
U_{v,2} &\approx \frac{\Gamma_2}{R_2}
\end{align}

Thrust was then calculated using the volume of the vortex ring $V_v$, the velocity of the vortex ring, $U_v$, and the vortex ring added mass, $C_{AM}$. 
\begin{align}
T_2 &= -\rho_w \, \frac{d}{dt} \left[ (1 + C_{AM}) \, U_{v,2} \, V_{v,2} \right]
\end{align}
By implementing this thrust term in equation (1), the forward velocity of the jellyfish could once again be iteratively determined. This model provided a more accurate representation of the jellyfish wake, but similar to the momentum model, it also depended heavily on the change in subumbrellar volume which determines the volume of fluid actuated as well as the exit velocity of the fluid.

The third, paddling, model was designed to avoid the dependency of swimming speed on change in subumbrellar volume as that volume may be a poor estimate of the fluid actuated in flatter, more oblate jellyfish like the ones tested here. In fact, for extremely oblate jellyfish, such as the \textit{Cassiopea} the water contained in the subumbrellar volume may increase during a contraction, rather than decrease (figure 4a). Due to this inversion, the $T_1$ and $T_2$ models were unable to be calculated for the \textit{Cassiopea}. In the third model the volume of water actuated was estimated instead from the motion of the jellyfish bell margin within a pulse. A swept volume of fluid based on the margin movement was used to estimate the actuated volume (figure 4b). This volume was used in place of $V_{sub}$ in equation (4) to calculate a new vortex volume $V_{v,3}$.

\begin{figure}[!h]
\centering\includegraphics[width=2.5in]{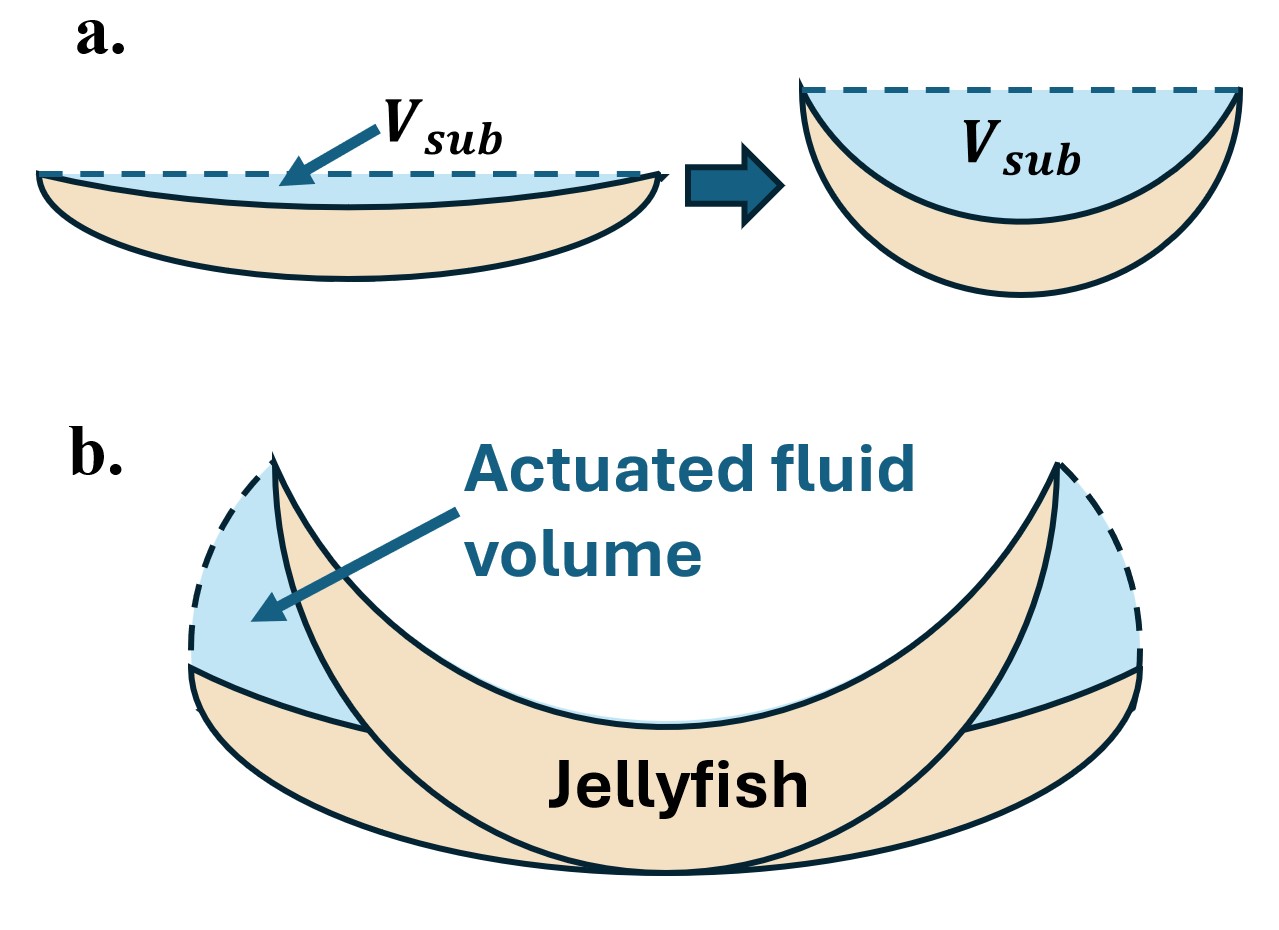}
\caption{\textbf{Fluid volume estimations during jellyfish contraction} (a) depicts the relaxed and contracted positions of an oblate jellyfish. The flat form factor of the jellyfish causes the contracted subumbrellar volume to be greater than the relaxed subumbrellar volume. This makes change in subumbrellar volume a poor estimate of the fluid actuated by the oblate jellyfish. (b) is a sketch of the actuated fluid estimation for the paddling thrust model, $T_3$. Here the swept volume of the jellyfish bell as it contracts is used to estimate the fluid volume actuated.}
\label{volume}
\end{figure}

This paddling thrust model used the same vortex thrust calculations as $T_2$, but the characteristic velocity used to calculate circulation was selected to be the speed of the jellyfish bell margin, rather than the jet velocity.
\begin{align}
\Gamma_3 &= \frac{1}{2} \int U_{\text{margin}}^2  \, dt
\end{align}

Plugging the new calculations for vortex volume and circulation into equations (5) and (7) led to new values of vortex radius ($R_3$) and vortex speed ($U_{v,3}$), respectively. This gives our final thrust equation, $T_3$.

\begin{align}
T_3 &= -\rho_w \, \frac{d}{dt} \left[ (1 + C_{AM}) \, U_{v,3} \, V_{v,3} \right]
\end{align}

All three analytical models were solved iteratively using a forward Euler scheme starting from a zero initial velocity. The swimming model using $T_1$ was also checked using a fourth order Runge-Kutta scheme before proceeding to ensure the forward Euler method had sufficient accuracy. For comparisons with the \textit{Aurelia} data, the model jellyfish was taken to be a hemiellipsoid with a relaxed fineness ratio of 0.38--comparable to the experimental test subjects. For comparison with the \textit{Cassiopea} data, a fineness ratio of 0.21 was used. Steady-state swimming speeds were taken as the pulse-averaged swimming speed once that average was within 1\% of the previous pulse. For each model the swimming frequencies were varied across the same range as the experiments (0.30-0.80 Hz). This range corresponded with the maximum Re numbers and minimum St numbers displayed in table I. Due to the positive buoyancy term, swimming speeds became negative for some frequencies, making only the upper bound of Re and lower bound of St relevant.

\begin{table}[h]
\caption{Maximum Reynolds number ($Re$) and minimum Strouhal number ($St$) for different models and experimental observations of two jellyfish species.}
\label{tab:re_st}
\begin{ruledtabular}
\begin{tabular}{l l c c}
Species & Model & Max $Re$ & Min $St$ \\
\hline
\textit{Aurelia aurita} 
    & Momentum, $T_1$ & 610  & 10.   \\
    & Jetting, $T_2$  & 860  & 6.6  \\
    & Paddling, $T_3$ & 1800 & 4.7  \\
    & Experimental       & 1700 & 5.0    \\
\textit{Cassiopea xamachana} 
    & Paddling, $T_3$ & 1500 & 5.9  \\
    & Experimental       & 1600 & 5.5  \\
\end{tabular}
\end{ruledtabular}
\end{table}

\subsection{Model input measurements}

In addition to measuring the swimming speed for each frequency tested, the measurements of the jellyfish swimming kinematics were used to extract estimates for the relevant model parameters. For the momentum and jetting vortex models  ($T_1$ and $T_2$) this included the percent diameter change during contractions, as well as the durations of the contraction and relaxation periods of the stroke for each frequency. The duration of the contraction and relaxation periods were determined from the measurements of the jellyfish widths.

\begin{figure}[!h]
\centering\includegraphics[width=3.5in]{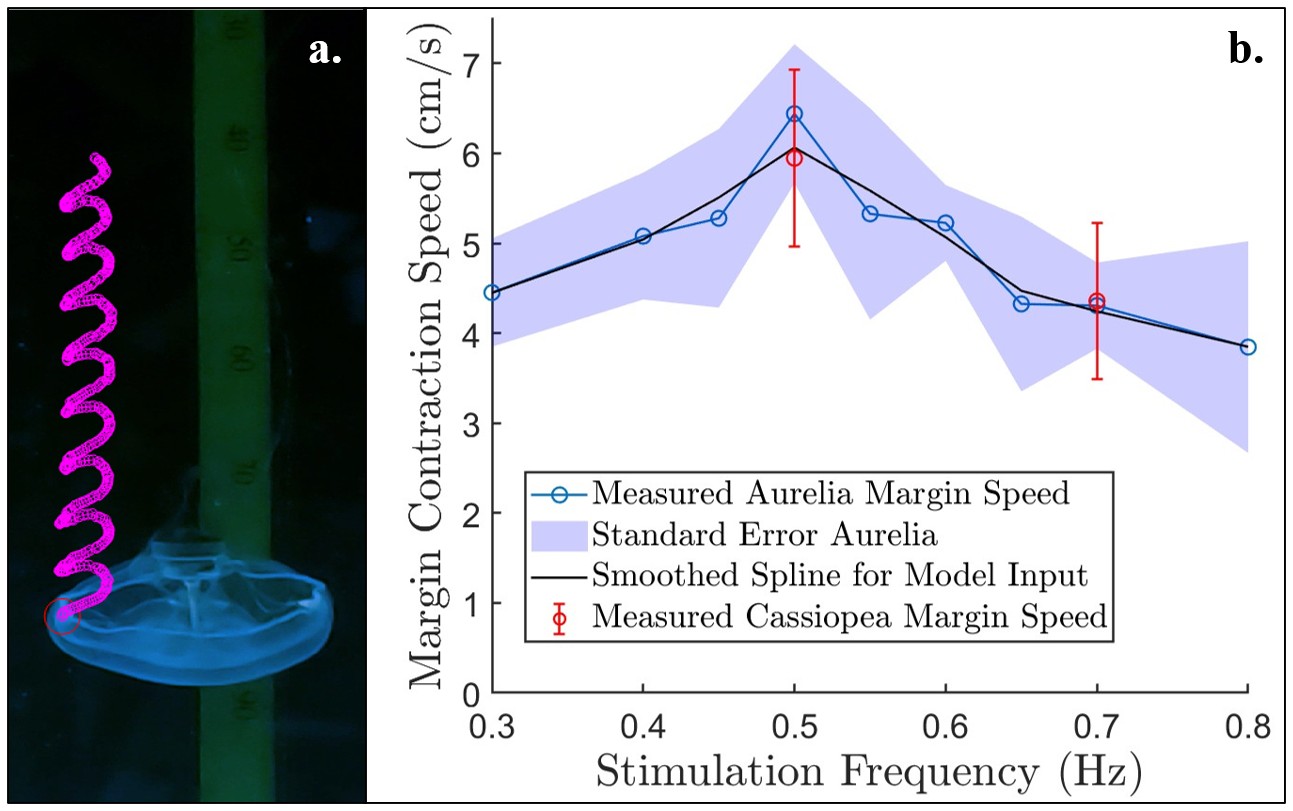}
\caption{\textbf{Margin tracking data extraction.} An image from matlab program DLTdv8a, used to track electrode position (a). \textit{Aurelia} margin contraction speeds extracted from the electrode position tracks are plotted for each frequency (b). Two frequencies for the \textit{Cassiopea} are also plotted along with the model input spline.}
\label{margin}
\end{figure}

For the paddling vortex model ($T_3$) the relevant parameters were the speed of bell margin as well as contraction and relaxation durations based on the motion of the margin rather than the overall jellyfish diameter. The Matlab program DLTdv8a was used to track the electrodes (figure 5a). This was used as a substitute for the bell margin as the electrodes were implanted 0.5 to 1.0 cm from the margin and were easier to track. Only video clips where the electrodes were normal to the camera view were used, determined by the distance from each electrode to the nearest edge of the jellyfish. DLTdv8a tracks of the jellyfish apex were subtracted from the bell margin tracks in order to convert from the lab-frame motion to the jellyfish-frame motion of the margin. The jellyfish-frame tracks of the electrodes were differentiated in order to determine the peak bell margin speed during contractions (figure 5b), as well as the contraction and relaxation durations. Using the peak speeds intentionally provided a slight overestimate of the electrode average speeds, in order to counter the underestimate of the bell margin speed caused by the inset of the electrodes from the true margin.

The margin speeds and contraction times of the \textit{Cassiopea} jellyfish were obfuscated by their more complex oral arms, as well as their less translucent bodies which prevented the electrodes from being tracked. Their similar form factor to the \textit{Aurelia}, however, suggested that the margin speed may be similar for both species. Video clips from two frequencies for the \textit{Cassiopea} were chosen and the margins were semi-manually tracked in DLTdv8a to determine the margin speeds. It was confirmed that for these two frequencies the margin speeds of the \textit{Aurelia} and \textit{Cassiopea} were similar. All model parameters were therefore based on the \textit{Aurelia} data.

\begin{figure*}[!t]
\centering\includegraphics[width=7in]{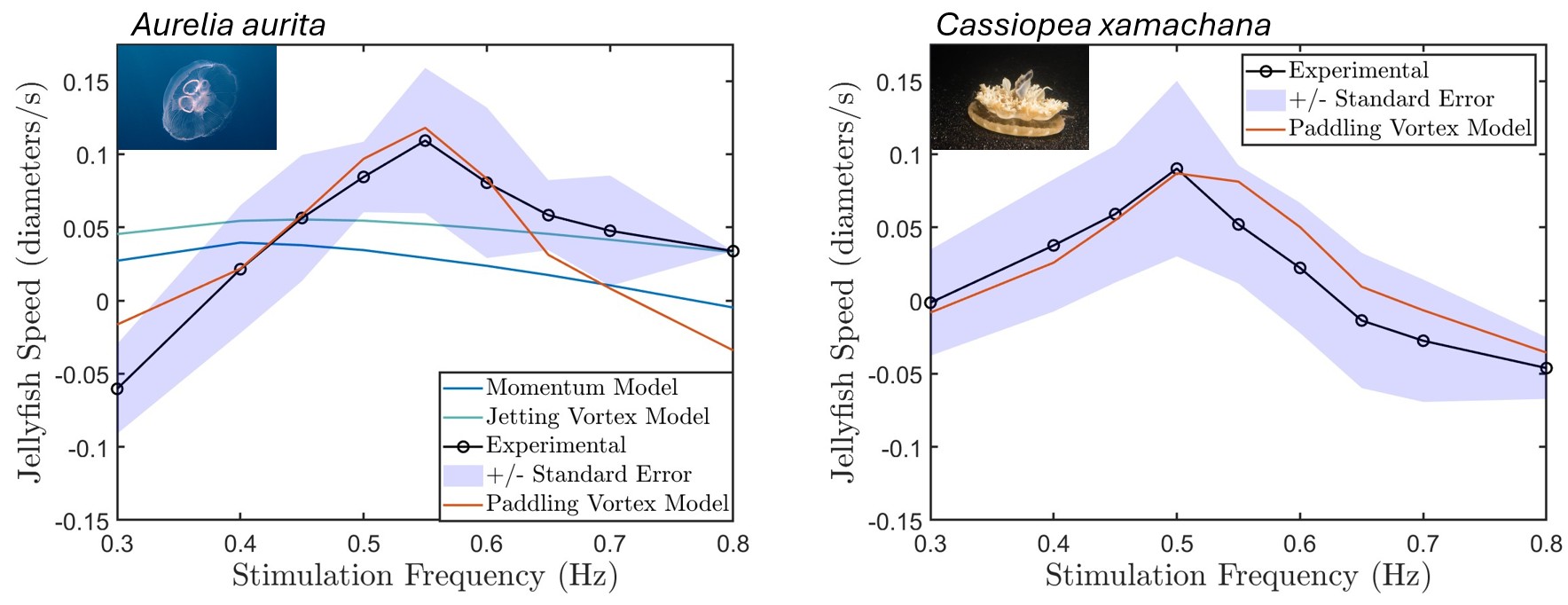}
\caption{\textbf{Experimental results and model predictions of jellyfish swimming speeds versus stimulation frequency.} The left plot shows the \textit{Aurelia aurita} experimental results as well as the predictions from the momentum, jetting vortex and paddling vortex models. The right plot shows the \textit{Cassiopea xamachana} results as well as the paddling vortex model prediction, as this was the only model that worked for the flatter \textit{Cassiopea} dimensions.}
\label{models}
\end{figure*}

\section{Results}

\subsection{Swimming performance}
The frequency variation experiments, displayed in figure 6, found that for both the \textit{Aurelia aurita} and \textit{Cassiopea xamachana} there was a non-monotonic dependence of swimming speed on stroke frequency. At low frequency (0.30 Hz), it was found that the jellyfish did not generate enough thrust to overcome the positive buoyancy of the microcontroller. For the \textit{Cassiopea} this resulted in holding position and zero forward speed, where as for the \textit{Aurelia} this resulted in floating up the tank, interpreted here as a negative speed. As the stimulation frequency was increased, the swimming speed for both species of jellyfish increased as well. The peak swimming speeds occurred at $0.55\pm0.05$ Hz and $0.50\pm0.05$ Hz for the \textit{Aurelia} and \textit{Cassiopea} respectively. Once the stroke frequency was increased past these peak values, the swimming speeds of the jellyfish decreased once again. For the \textit{Cassiopea} this resulted in negative speeds (floating up the tank) at the highest stimulation frequencies. For the \textit{Aurelia} the jellyfish began responding intermittently to the stimulation at frequencies above 0.70 Hz, leading to a less reliable swimming speed value at 0.80 Hz.

\subsection{Analytical model comparisons}

Figure 6 also plots the cycle-averaged swimming speed versus stroke frequency predicted by the analytical models. For the \textit{Aurelia} all three models predict an initial increase in swimming speed with stroke frequency up to a peak after which point continued increases in stroke frequency lead to decreased swimming speeds. However, the percent change in swimming speed based on frequency is significantly underestimated by both the momentum and jetting vortex models, and slightly overestimated by the paddling vortex model. The experimental \textit{Aurelia} data shows that the peak swimming speed is an 80\% increase from the first positive swimming speed. This is best matched by the paddling model which predicts an 83\% increase from the first positive swimming speed, as opposed to the momentum model which predicts a 31\% increase and the jetting model which predicts an 18\% increase. Only the paddling vortex model accurately predicts the frequency at which the maximum swimming speed occurs, while the momentum and jetting models underestimate this frequency. The paddling model also better estimates the swimming speed values with a normalized root mean square error of 29\% as opposed to the jet model with 40\% and the momentum model with 48\%.

\begin{figure}[!h]
\centering\includegraphics[width=3.2in]{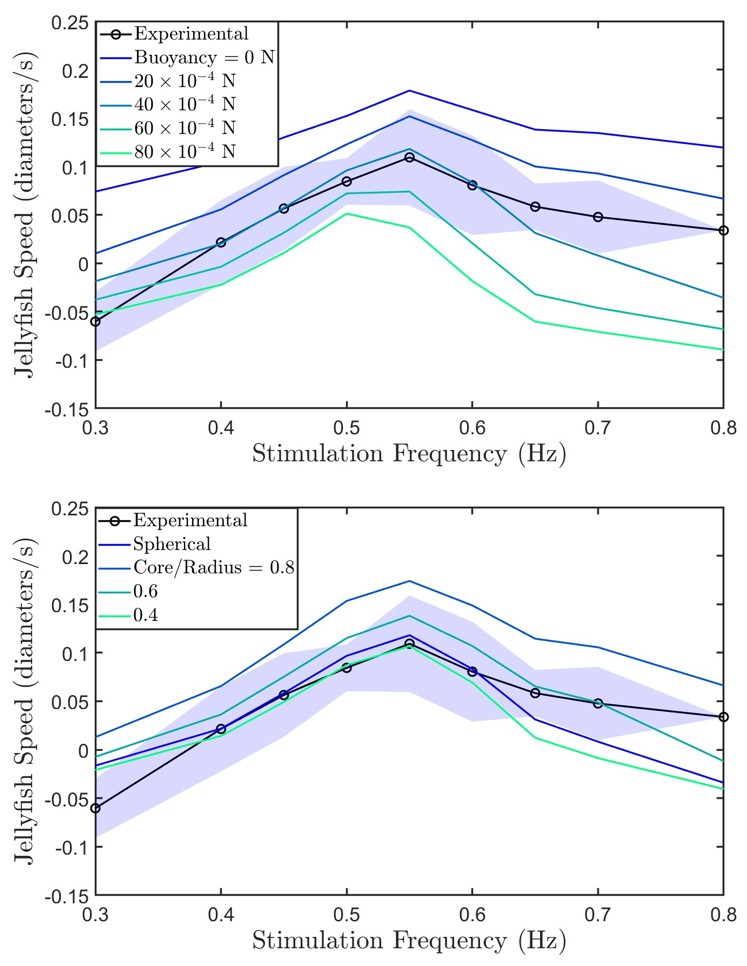}
\caption{\textbf{Parameters sweeps of buoyancy and vortex shape for the scyphozoan, paddling model.} The upper plot depicts the experimental \textit{Aurelia} data with paddling model predictions for a range of buoyancies. As buoyancy increases the swimming speeds decrease and the peak becomes more pronounced. The lower plot depicts the same experimental results with the paddling model for a range of vortex shapes. The experimental data is most accurately predicted using either the spherical shape or a flat (low core to radius ratio) torus.}
\label{parameters}
\end{figure}

For the \textit{Cassiopea}, only the paddling vortex model was able to be applied. Their extremely flat form factor prevents the change in subumbrellar volume from being calculated when using their dimensions. When applied to the \textit{Cassiopea} the paddling vortex model still predicts the frequency at which the peak swimming speed occurs, despite being based in the \textit{Aurelia} kinematics. Additionally it has a normalized root mean square error of only 20\%.

Figure 7 provides a parameter sweep of the paddling vortex model for two variables that were not experimentally determined, the buoyancy and the vortex shape. Here they are compared only against the \textit{Aurelia} experimental data as the effect of the variable is similar for both model species. As can easily be predicted, an increase in buoyancy leads to a decrease in the downward swimming speed of the jellyfish. Increased buoyancy also leads to a more prominent peak in swimming speed versus stimulation frequency. This is most likely due to the decreased effects of drag at the lower speeds achieved with higher buoyancy. Modeling the vortex shape as a torus rather than a sphere initially increases the predicted swimming speeds. As the torus becomes more flattened (core to radius ratio decreases), the predicted swimming speeds decrease, most likely due to the larger vortex radius.

\section{Discussion}
Experimental measurements of scyphozoan jellyfish swimming speeds indicated that stroke frequency had a significant and non-monotonic impact on swimming speed. The relationship between stroke frequency and swimming speed was found to be similar for both the \textit{Aurelia aurita} and \textit{Cassiopea xamachana} despite their different natural frequencies. This suggests that the frequency at which the peak swimming speed occurs is determined from the body shape and the hydrodynamics rather than any biological factor, as further indicated by the analytical modeling. This also reinforces the hypothesis that natural stroke frequencies are driven by other functions such as feeding, rather than swimming.

Many variables are linked in animal behavior and can be difficult to disentangle in experiments. Biohybrid robotic control can be used to isolate such variables, in this case the stroke frequency of jellyfish. The effects of those variables can then be investigated in a systematic way, reaching outside the bounds of what we can observe in nature. As bioinspired robotics become more prevelant, tests such as these are important to disambiguate what parts of biological motion are useful to replicate. 

Previously developed models of jellyfish swimming speeds were unable to predict the relationship between swimming speed and stroke frequency in the scyphozoan jellyfish tested in these experiments. These previous models were developed for hydrozoan jellyfish and were highly dependent on the subumbrellar volume, an unreliable variable in oblate, scyphozoan jellyfish.

The new model developed for scyphozoan jellyfish swimming was able to predict both the magnitudes of the jellyfish swimming speeds within a fractional error for an intermediate range of frequencies, as well as the relationship between speed and stroke frequency in both species of jellyfish tested. This indicates that the relevant variable for estimating the vortex circulation and therefore the vortex velocity and thrust in these paddling organisms is the margin speed. The change in margin speed with frequency in turn, is dependent on the kinematics of the jellyfish bodies. This model does not explore the vortex interactions or the passive energy recapture, which implies that neither phenomenon is necesary to quantify in order to determine the swimming speed and stroke frequency relationship.

In addition to furthering our understanding of scyphozoan jellyfish swimming, this model may be useful for bioinspired robots emulating this propulsive method. In robotic platforms, the margin velocity will most likely be a known input parameter, making it straightforward to implement this model for speed predictions and control. Furthermore, biohybrid jellyfish have been proposed as a new method to enable ocean sensing \citep{xu_field_2020}. The testing and analytical modeling of scyphozoan jellyfish described in this paper could be used to implement speed control for biohybrid jellyfish sensors. These biohybrid jellyfish could leverage the dependence of jellyfish swimming speed on frequency, even if the natural animals do not.

\begin{acknowledgments}
This work was supported by the National Science Foundation under Grant 2529198 and the NSF Graduate Research Fellowship DGE-2146755, as well as by the Office of Naval Research under Grant N000142412478. The authors would like to thank Alex Hoover, Simon Anuszczyk, Sean Colin, Brad Gemmell, and John Costello for their insights. The authors would also like to thank Cabrillo Marine Aquarium and Aquarium of the Pacific for providing jellyfish used in these experiments.
\end{acknowledgments}

\bibliography{apssamp}

@article{costello_hydrodynamics_2021,
	title = {The {Hydrodynamics} of {Jellyfish} {Swimming}},
	volume = {13},
	issn = {1941-1405, 1941-0611},
	url = {https://www.annualreviews.org/content/journals/10.1146/annurev-marine-031120-091442},
	doi = {10.1146/annurev-marine-031120-091442},
	number = {Volume 13, 2021},
	urldate = {2025-07-07},
	journal = {Annual Review of Marine Science},
	publisher = {Annual Reviews},
	author = {Costello, John H. and Colin, Sean P. and Dabiri, John O. and Gemmell, Brad J. and Lucas, Kelsey N. and Sutherland, Kelly R.},
	month = jan,
	year = {2021},
	pages = {375--396},
}

@article{bainbridge_speed_1958,
	title = {The {Speed} of {Swimming} of {Fish} as {Related} to {Size} and to the {Frequency} and {Amplitude} of the {Tail} {Beat}},
	volume = {35},
	issn = {0022-0949},
	url = {https://doi.org/10.1242/jeb.35.1.109},
	doi = {10.1242/jeb.35.1.109},
	number = {1},
	urldate = {2025-07-07},
	journal = {Journal of Experimental Biology},
	author = {Bainbridge, Richard},
	month = mar,
	year = {1958},
	pages = {109--133},
}

@article{plaut_critical_2001,
	title = {Critical swimming speed: its ecological relevance},
	volume = {131},
	issn = {1095-6433},
	shorttitle = {Critical swimming speed},
	url = {https://www.sciencedirect.com/science/article/pii/S1095643301004627},
	doi = {10.1016/S1095-6433(01)00462-7},
	number = {1},
	urldate = {2025-07-07},
	journal = {Comparative Biochemistry and Physiology Part A: Molecular \& Integrative Physiology},
	author = {Plaut, Itai},
	month = dec,
	year = {2001},
	pages = {41--50},
}

@article{videler_fish_1991,
	title = {Fish swimming stride by stride: speed limits and endurance},
	volume = {1},
	issn = {1573-5184},
	shorttitle = {Fish swimming stride by stride},
	url = {https://doi.org/10.1007/BF00042660},
	doi = {10.1007/BF00042660},
	number = {1},
	urldate = {2025-07-07},
	journal = {Reviews in Fish Biology and Fisheries},
	author = {Videler, J. J. and Wardle, C. S.},
	month = sep,
	year = {1991},
	pages = {23--40},
}

@article{dabiri_fast-swimming_2006,
	title = {Fast-swimming hydromedusae exploit velar kinematics to form an optimal vortex wake},
	volume = {209},
	issn = {0022-0949},
	url = {https://doi.org/10.1242/jeb.02242},
	doi = {10.1242/jeb.02242},
	number = {11},
	urldate = {2025-07-07},
	journal = {Journal of Experimental Biology},
	author = {Dabiri, John O. and Colin, Sean P. and Costello, John H.},
	month = jun,
	year = {2006},
	pages = {2025--2033},
}

@article{daniel_mechanics_1983,
	title = {Mechanics and energetics of medusan jet propulsion},
	volume = {61},
	issn = {0008-4301},
	url = {https://cdnsciencepub.com/doi/10.1139/z83-190},
	doi = {10.1139/z83-190},
	number = {6},
	urldate = {2025-07-07},
	journal = {Canadian Journal of Zoology},
	publisher = {NRC Research Press},
	author = {Daniel, Thomas L.},
	month = jun,
	year = {1983},
	pages = {1406--1420},
}

@article{xu_low-power_2020,
	title = {Low-power microelectronics embedded in live jellyfish enhance propulsion},
	volume = {6},
	url = {https://www.science.org/doi/10.1126/sciadv.aaz3194},
	doi = {10.1126/sciadv.aaz3194},
	number = {5},
	urldate = {2025-07-07},
	journal = {Science Advances},
	publisher = {American Association for the Advancement of Science},
	author = {Xu, Nicole W. and Dabiri, John O.},
	month = jan,
	year = {2020},
	pages = {eaaz3194},
}

@incollection{lauder_hydrodynamics_2005,
	series = {Fish {Biomechanics}},
	title = {Hydrodynamics of {Undulatory} {Propulsion}},
	volume = {23},
	url = {https://www.sciencedirect.com/science/article/pii/S154650980523011X},
	doi = {10.1016/S1546-5098(05)23011-X},
	urldate = {2025-07-07},
	booktitle = {Fish {Physiology}},
	publisher = {Academic Press},
	author = {Lauder, George V. and Tytell, Eric D.},
	month = jan,
	year = {2005},
	pages = {425--468},
}

@article{xu_field_2020,
	title = {Field {Testing} of {Biohybrid} {Robotic} {Jellyfish} to {Demonstrate} {Enhanced} {Swimming} {Speeds}},
	volume = {5},
	copyright = {http://creativecommons.org/licenses/by/3.0/},
	issn = {2313-7673},
	url = {https://www.mdpi.com/2313-7673/5/4/64},
	doi = {10.3390/biomimetics5040064},
	number = {4},
	urldate = {2025-07-17},
	journal = {Biomimetics},
	publisher = {Multidisciplinary Digital Publishing Institute},
	author = {Xu, Nicole W. and Townsend, James P. and Costello, John H. and Colin, Sean P. and Gemmell, Brad J. and Dabiri, John O.},
	month = dec,
	year = {2020},
	note = {Number: 4},
	pages = {64},
}

@article{sanchez-rodriguez_scaling_2023,
	title = {Scaling the tail beat frequency and swimming speed in underwater undulatory swimming},
	volume = {14},
	copyright = {2023 The Author(s)},
	issn = {2041-1723},
	url = {https://www.nature.com/articles/s41467-023-41368-6},
	doi = {10.1038/s41467-023-41368-6},
	number = {1},
	urldate = {2025-07-17},
	journal = {Nature Communications},
	publisher = {Nature Publishing Group},
	author = {Sánchez-Rodríguez, Jesús and Raufaste, Christophe and Argentina, Médéric},
	month = sep,
	year = {2023},
	pages = {5569},
}

@article{chia_locomotion_1984,
	title = {Locomotion of marine invertebrate larvae: a review},
	volume = {62},
	issn = {0008-4301},
	shorttitle = {Locomotion of marine invertebrate larvae},
	url = {https://cdnsciencepub.com/doi/abs/10.1139/z84-176},
	doi = {10.1139/z84-176},
	number = {7},
	urldate = {2025-07-17},
	journal = {Canadian Journal of Zoology},
	publisher = {NRC Research Press},
	author = {Chia, Fu-Shiang and Buckland-Nicks, John and Young, Craig M.},
	month = jul,
	year = {1984},
	pages = {1205--1222},
}

@article{gemmell_cool_2021,
	title = {Cool your jets: biological jet propulsion in marine invertebrates},
	volume = {224},
	issn = {0022-0949},
	shorttitle = {Cool your jets},
	url = {https://doi.org/10.1242/jeb.222083},
	doi = {10.1242/jeb.222083},
	number = {12},
	urldate = {2025-07-17},
	journal = {Journal of Experimental Biology},
	author = {Gemmell, Brad J. and Dabiri, John O. and Colin, Sean P. and Costello, John H. and Townsend, James P. and Sutherland, Kelly R.},
	month = jun,
	year = {2021},
	pages = {jeb222083},
}

@article{gemmell_passive_2013,
	title = {Passive energy recapture in jellyfish contributes to propulsive advantage over other metazoans},
	volume = {110},
	url = {https://www.pnas.org/doi/abs/10.1073/pnas.1306983110},
	doi = {10.1073/pnas.1306983110},
	number = {44},
	urldate = {2025-07-17},
	journal = {Proceedings of the National Academy of Sciences},
	publisher = {Proceedings of the National Academy of Sciences},
	author = {Gemmell, Brad J. and Costello, John H. and Colin, Sean P. and Stewart, Colin J. and Dabiri, John O. and Tafti, Danesh and Priya, Shashank},
	month = oct,
	year = {2013},
	pages = {17904--17909},
}

@article{villanueva_biomimetic_2011,
	title = {A biomimetic robotic jellyfish ({Robojelly}) actuated by shape memory alloy composite actuators},
	volume = {6},
	issn = {1748-3190},
	url = {https://dx.doi.org/10.1088/1748-3182/6/3/036004},
	doi = {10.1088/1748-3182/6/3/036004},
	number = {3},
	urldate = {2025-07-17},
	journal = {Bioinspiration \& Biomimetics},
	author = {Villanueva, Alex and Smith, Colin and Priya, Shashank},
	month = aug,
	year = {2011},
	pages = {036004},
}

@article{cheng_untethered_2018,
	title = {Untethered soft robotic jellyfish},
	volume = {28},
	issn = {0964-1726},
	url = {https://dx.doi.org/10.1088/1361-665X/aaed4f},
	doi = {10.1088/1361-665X/aaed4f},
	number = {1},
	urldate = {2025-07-17},
	journal = {Smart Materials and Structures},
	publisher = {IOP Publishing},
	author = {Cheng, Tingyu and Li, Guori and Liang, Yiming and Zhang, Mingqi and Liu, Bangyuan and Wong, Tuck-Whye and Forman, Jack and Chen, Mianhong and Wang, Guanyun and Tao, Ye and Li, Tiefeng},
	month = nov,
	year = {2018},
	pages = {015019},
}

@article{wang_versatile_2023,
	title = {A versatile jellyfish-like robotic platform for effective underwater propulsion and manipulation},
	volume = {9},
	url = {https://www.science.org/doi/full/10.1126/sciadv.adg0292},
	doi = {10.1126/sciadv.adg0292},
	number = {15},
	urldate = {2025-07-17},
	journal = {Science Advances},
	publisher = {American Association for the Advancement of Science},
	author = {Wang, Tianlu and Joo, Hyeong-Joon and Song, Shanyuan and Hu, Wenqi and Keplinger, Christoph and Sitti, Metin},
	month = apr,
	year = {2023},
	pages = {eadg0292},
}

@article{xu_ethics_2025,
	title = {Ethics of biohybrid robotics and invertebrate research: biohybrid robotic jellyfish as a case study},
	volume = {20},
	issn = {1748-3190},
	shorttitle = {Ethics of biohybrid robotics and invertebrate research},
	url = {https://dx.doi.org/10.1088/1748-3190/adc0d4},
	doi = {10.1088/1748-3190/adc0d4},
	number = {3},
	urldate = {2025-07-17},
	journal = {Bioinspiration \& Biomimetics},
	publisher = {IOP Publishing},
	author = {Xu, Nicole W and Lenczewska, Olga and Wieten, Sarah E and Federico, Carole A and Dabiri, John O},
	month = apr,
	year = {2025},
	pages = {033001},
}

@article{gladfelter_comparative_1973,
	title = {A comparative analysis of the locomotory systems of medusoid {Cnidaria}},
	volume = {25},
	copyright = {http://www.springer.com/tdm},
	issn = {0017-9957, 1438-3888},
	url = {https://link.springer.com/10.1007/BF01611199},
	doi = {10.1007/BF01611199},
	number = {2-3},
	urldate = {2026-03-20},
	journal = {Helgoländer Wissenschaftliche Meeresuntersuchungen},
	author = {Gladfelter, W. G.},
	month = sep,
	year = {1973},
	pages = {228--272},
}

@misc{seanet_hydrozoa,
  author = {{Stanford SeaNet}},
  title = {Class Hydrozoa},
  year = {n.d.},
  url = {https://seanet.stanford.edu/Hydrozoa},
  note = {Accessed March 20, 2026}
}

@misc{jellyfishwarehouse_crosshair,
  author = {{Jellyfish Warehouse}},
  title = {Crosshair Hydromedusa (Bundle of Three)},
  year = {n.d.},
  url = {https://jellyfishwarehouse.com/products/crosshair-hydromedusa-bundle-of-three},
  note = {Accessed March 20, 2026}
}

@misc{noauthor_european_nodate,
  author = {{European Marine Life}},
  title = {Scyphozoan and Staurozoan Cnidarians Gallery (True Jellyfish and Staurozoans)},
  year = {n.d.},
  url = {https://www.european-marine-life.org/05/scyphozoan-gallery.php},
  note = {Accessed March 20, 2026}
}

@article{association_caught_2017,
	chapter = {Science},
	title = {Caught napping: snoozing jellyfish prove a brain isn't necessary for sleep},
	issn = {0261-3077},
	shorttitle = {Caught napping},
	url = {https://www.theguardian.com/science/2017/sep/21/caught-napping-snoozing-jellyfish-prove-a-brain-isnt-necessary-for-sleep},
	urldate = {2026-03-20},
	journal = {The Guardian},
	author = {Association, Press},
	month = sep,
	year = {2017},
}

@article{nath_jellyfish_2017,
	title = {The jellyfish {Cassiopea} exhibits a sleep-like state},
	volume = {27},
	issn = {0960-9822},
	url = {https://pmc.ncbi.nlm.nih.gov/articles/PMC5653286/},
	doi = {10.1016/j.cub.2017.08.014},
	number = {19},
	urldate = {2026-03-20},
	journal = {Current biology : CB},
	author = {Nath, Ravi D. and Bedbrook, Claire N. and Abrams, Michael J. and Basinger, Ty and Bois, Justin S. and Prober, David A. and Sternberg, Paul W. and Gradinaru, Viviana and Goentoro, Lea},
	month = oct,
	year = {2017},
	pages = {2984--2990.e3},
}

@article{anuszczyk_increasing_2025,
	title = {Increasing the {Reliability} and {Versatility} of {Jellyfish} {Biohybrid} {Vehicles} via {Species} {Selection} and {Rhopalia} {Removal}},
	volume = {10},
	issn = {2313-7673},
	url = {https://www.mdpi.com/2313-7673/10/12/810},
	doi = {10.3390/biomimetics10120810},
	number = {12},
	urldate = {2026-03-20},
	journal = {Biomimetics},
	author = {Anuszczyk, Simon R. and Yoder, Noa and Costello, John H. and Dabiri, John O. and Gemmell, Brad J. and Rutledge, Kelsi M. and Colin, Sean P.},
	month = dec,
	year = {2025},
	pages = {810},
}

@article{feitl_functional_2009,
	title = {Functional {Morphology} and {Fluid} {Interactions} {During} {Early} {Development} of the {Scyphomedusa} {Aurelia} aurita},
	volume = {217},
	doi = {10.1086/BBLv217n3p283},
	journal = {The Biological bulletin},
	author = {Feitl, K and Millett, A.F. and Colin, Sean and Dabiri, J and Costello, John},
	month = dec,
	year = {2009},
	pages = {283--91},
}

@book{sighard_f_hoerner_fluid-dynamic_1965,
  author    = {Hoerner, S. F.},
  title     = {Fluid-Dynamic Drag},
  year      = {1965},
  publisher = {Hoerner Fluid Dynamics},
  address   = {Midland Park, NJ}
}

@article{dabiri_flow_2005,
	title = {Flow patterns generated by oblate medusan jellyfish: field measurements and laboratory analyses},
	volume = {208},
	issn = {0022-0949},
	shorttitle = {Flow patterns generated by oblate medusan jellyfish},
	doi = {10.1242/jeb.01519},
	number = {Pt 7},
	journal = {The Journal of Experimental Biology},
	author = {Dabiri, John O. and Colin, Sean P. and Costello, John H. and Gharib, Morteza},
	month = apr,
	year = {2005},
	pages = {1257--1265},
}

@article{mchenry_ontogenetic_2003,
	title = {The ontogenetic scaling of hydrodynamics and swimming performance in jellyfish (\textit{{Aurelia} aurita})},
	url = {https://sci-hub.ru/10.1242/jeb.00649},
	doi = {10.1242/jeb.00649},
	urldate = {2026-04-10},
	journal = {Journal of Experimental Biology},
	author = {McHenry, Matthew J. and Jed, Jason},
	year = {2003},
}

@article{herschlag_reynolds_2011,
	title = {Reynolds number limits for jet propulsion: {A} numerical study of simplified jellyfish},
	volume = {285},
	issn = {0022-5193},
	shorttitle = {Reynolds number limits for jet propulsion},
	url = {https://www.sciencedirect.com/science/article/pii/S0022519311002876},
	doi = {10.1016/j.jtbi.2011.05.035},
	number = {1},
	urldate = {2026-04-10},
	journal = {Journal of Theoretical Biology},
	author = {Herschlag, Gregory and Miller, Laura},
	month = sep,
	year = {2011},
	pages = {84--95},
}

@article{taylor_flying_2003,
	title = {Flying and swimming animals cruise at a {Strouhal} number tuned for high power efficiency},
	volume = {425},
	copyright = {2003 Macmillan Magazines Ltd.},
	issn = {1476-4687},
	url = {https://www.nature.com/articles/nature02000},
	doi = {10.1038/nature02000},
	number = {6959},
	urldate = {2026-04-14},
	journal = {Nature},
	publisher = {Nature Publishing Group},
	author = {Taylor, Graham K. and Nudds, Robert L. and Thomas, Adrian L. R.},
	month = oct,
	year = {2003},
	pages = {707--711},
}

\end{document}